\documentclass[12pt,a4paper]{iopart}
\usepackage{iopams}  
\usepackage{graphicx}  
\begin{document}

\title{
Near-thermal equilibrium with Tsallis distributions \\
in heavy ion collisions
}

\author{
J. Cleymans$^a$,
G. Hamar$^b$, 
P. Levai$^b$,
S. Wheaton$^a$,
}

\address{
$^a$ UCT-CERN Research Centre and Department of Physics, University of Cape Town, 
Rondebosch 7701, Cape Town, South Africa\\
$^b$ MTA KFKI RMKI, Research Insitute for Particle and Nuclear Physics, \\
Po.B. 49, Budapest, 1525, Hungary\\
}
\ead{jean.cleymans@uct.ac.za}
\begin{abstract}
Hadron yields in high energy heavy ion collisions have been 
fitted 
with thermal models using standard
(extensive)
statistical
distributions. These models give insight into 
the freeze-out conditions at varying beam energies 
and lead to a systematic 
consistent picture of freeze-out conditions at all beam energies. 
In this paper we investigate changes to this analysis 
when the statistical distributions are replaced by 
non-extensive
Tsallis distributions 
for hadrons. We investigate the 
particle yields
at SPS and RHIC energies and obtain better
fits with smaller $\chi^2$ for the same hadron data, as applied earlier
in the thermal fits for SPS energies but not for RHIC energies. 
\end{abstract}


\section{Introduction}

After many years of investigating  hadron-hadron and heavy ion collisions, 
the study of hadron production remains an active and 
 important field of research.
The lack of detailed knowledge of the microscopic mechanisms
has led to the use of many different models, often from
completely opposite directions. Thermal models,
 based on statistical weights for produced
hadrons~\cite{Fermi50,Pomer51,heisenberg,Haged65}, 
 are very successful in describing particle yields
at different beam energies~\cite{wheaton,strange,andronic,becattini}, especially in heavy ion collisions.
These models assume the formation of a system which is in  
thermal and chemical equilibrium in the hadronic phase and is
characterised by a set of thermodynamic  variables for the
hadronic phase. The deconfined period of the time evolution
dominated by quarks and gluons remains hidden: full equilibration generally
washes out and destroys large amounts of information about the early 
deconfined phase.
The success of statistical models implies the loss of such
information,  at least for certain properties, during hadronization.
It is a basic question as to which ones survive the hadronization
and behave as messengers from the early (quark dominated) stages,
especially if these are strongly interacting stages.

 In the case of full thermal and chemical equilibrium, relativistic
statistical distributions can be used, leading to exponential spectra for
the transverse momentum distribution of hadrons. 
On the other hand, experimental data at SPS and
RHIC energies display non-exponential behaviours at high $p_T$. One
explanation of this deviation is connected to the power-like
hadron spectra obtained from perturbative QCD descriptions: the
hadron yield from quark and gluon fragmentation overwhelms the thermal
(exponential) hadron production. However, this overlap is not trivial.
One can assume the appearance of near-thermal hadron distributions,
which is similar to the thermal distribution at lower $p_T$, but
it has a non-exponential tail at higher $p_T$. 
A stationary distribution of
strongly interacting hadron gas in a finite volume can be characterized
by such a distribution (or strongly interacting quark matter), which
will hadronize into hadron matter. 
Tsallis distributions satisfy such criteria \cite{tsallis,tsallis3}.
In the next Section we will review the Tsallis distribution and emphasize the
properties most relevant to particle yields. 
\section{Tsallis Distribution for Particle Multiplicities.}
\subsection{Relation  between the Boltzmann and Tsallis distributions}
\begin{figure}
\begin{center}
\includegraphics[width=0.47\textwidth,height=10cm]{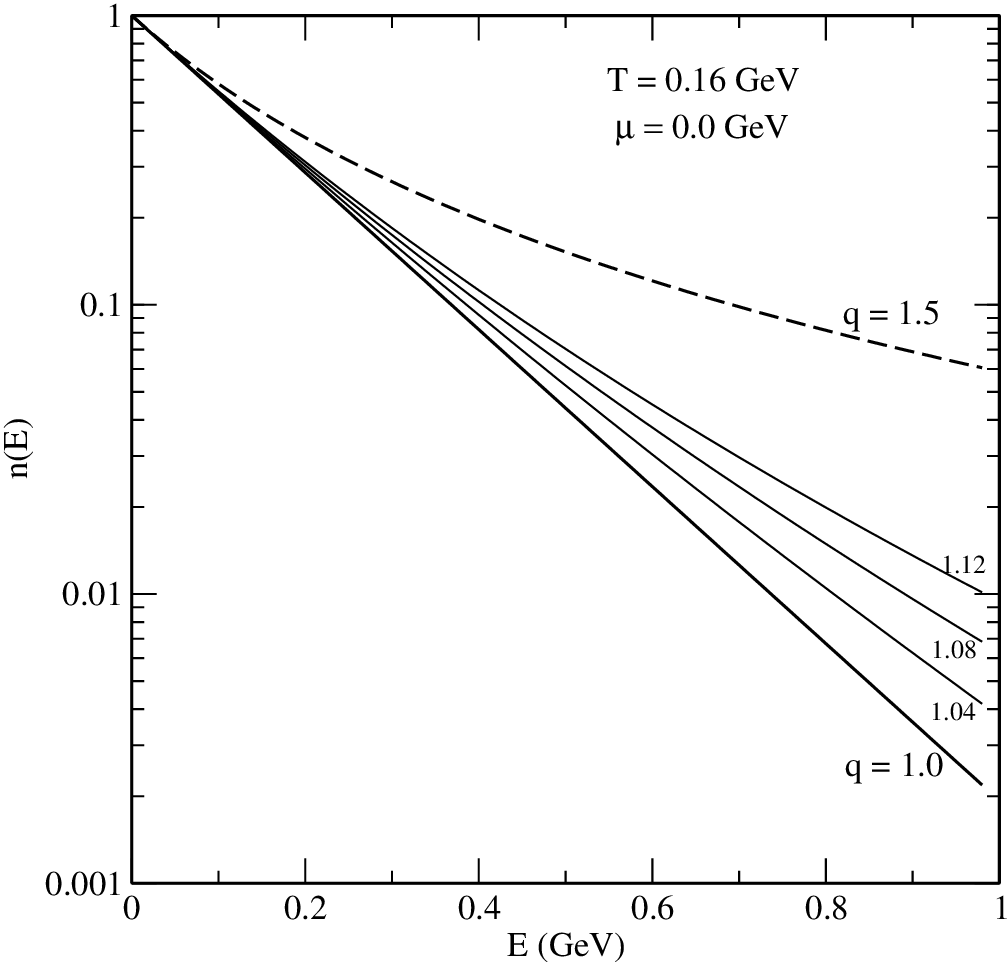}
\includegraphics[width=0.47\textwidth,height=10cm]{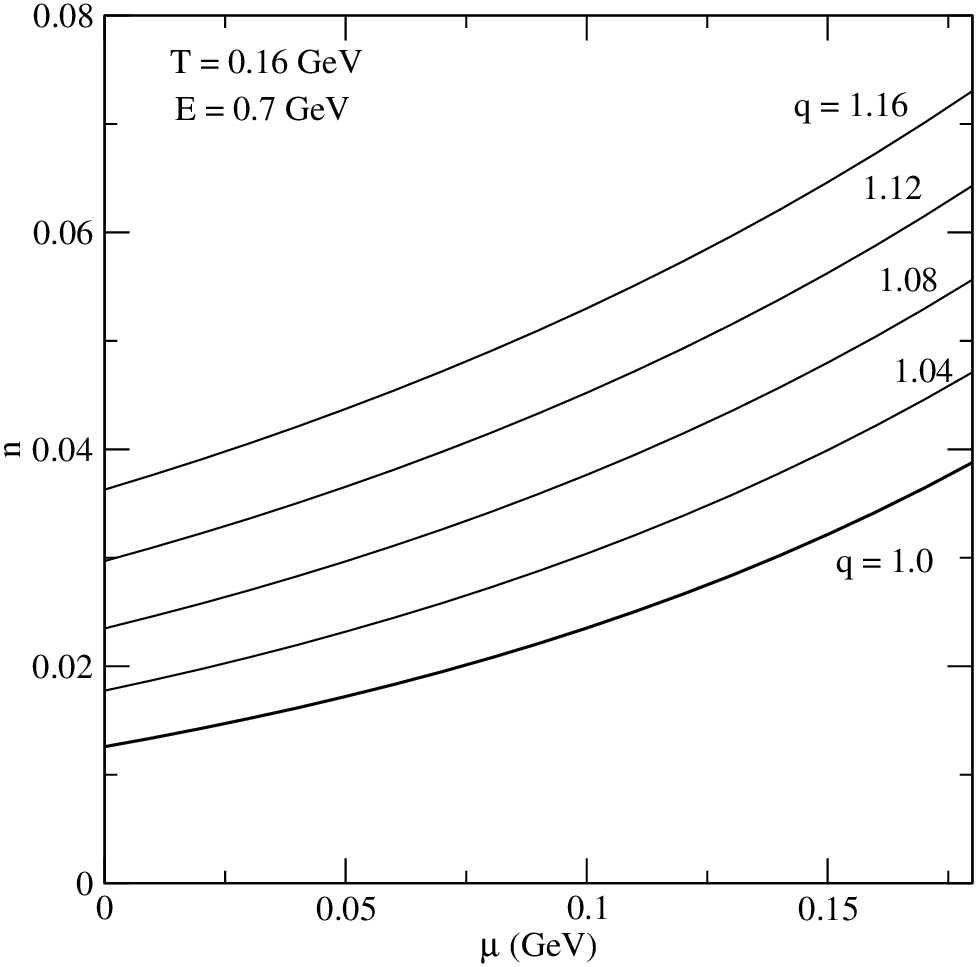}
\caption{Comparison between the Boltzmann and Tsallis distributions. 
The figure on the left compares the two as a 
function of the energy $E$, keeping 
the temperature and chemical potential fixed, for various
values of the Tsallis parameter $q$. 
The value $q = 1.5$ is the maximum value that still leads 
to a convergent integral
in Eq. \ref{tsallis_number}.
chemical potential $\mu$, keeping 
the temperature and the energy fixed. }
\label{tsallis_boltzmann}
\end{center}
\end{figure}

Neglecting quantum statistics,
the entropy 
of a particle of species $i$
 is given by~\cite{degroot}
\begin{equation}
S_i = V\int{d^3p\over (2\pi)^3} (n^B_i- n^B_i\ln n^B_i) ,
\label{entropy_boltzmann}
\end{equation}
where the mean occupation 
numbers, $n^B_i$,  are given by
\begin{equation}
n^B_i(E) \equiv g_i\exp\left(-{E_i-\mu_i\over T}\right)  .
\label{boltzmann}
\end{equation}
with $g_i$ being the degeneracy factor of particle $i$.
The total number of particles of species $i$ is given by an integral over 
phase space of eq. (\ref{boltzmann}):
\begin{equation}
N_i = V\int {d^3p\over (2\pi)^3} n^B_i(E) .
\end{equation}
The transition to the Tsallis distribution makes use of the following substitutions~\cite{tsallis}
\begin{eqnarray}
\ln(x) &\rightarrow& \ln_q (x)\equiv {x^{1-q}-1\over 1-q},\label{suba} \\
\exp(x) &\rightarrow&\exp_q(x)\equiv [1+(1-q)x]^{1\over 1-q} , \label{subb}
\end{eqnarray}
which leads to the standard result~\cite{tsallis,tsallis3}
\begin{equation}
n^T_i(E) = g_i\left[ 1 + (q-1) {E_i-\mu_i\over T}\right]^{-{1\over q-1}} ,
\label{tsallis}
\end{equation}
which is usually referred to as the Tsallis 
distribution \cite{tsallis,tsallis3}.  
As these number densities  are not normalized, 
we do not use the normalized $q$-probabilities which have been 
proposed in Ref.~\cite{tsallis3}.  
In the limit where 
$q\rightarrow 1$ this becomes the Boltzmann distribution:
\begin{equation}
\lim_{q\rightarrow 1} n^T_i(E) = n^B_i(E) .
\end{equation}
The particle number is now given by
\begin{equation}
N_i = V\int {d^3p\over (2\pi)^3} (n^T_i(E))^q .
\label{tsallis_number}
\end{equation}
%
Note that $q = 1.5$ is the maximum value that still leads 
to a convergent integral
in eq. (\ref{tsallis_number}).
A derivation of the Tsallis distribution, based on the Boltzmann equation,
has been given in Ref.~\cite{biro2}.
A comparison between the two distributions   is shown in Fig.~(\ref{tsallis_boltzmann}), where it 
can be seen that, at fixed values of $T$ and $\mu$,  the Tsallis distribution is 
always larger than the Boltzmann one if $q>1$. Taking 
into account the large $p_T$ results for 
particle production we will only consider this possibility in this paper.
As a consequence, in order to keep
the particle yields the same, the Tsallis distribution always leads to
smaller values of $T$ for the same set of particle yields.
The dependence on the chemical potential is also  illustrated on the right of  
Fig.~\ref{tsallis_boltzmann} for a fixed temperature $T$ and a fixed
 energy $E$. As one can see, the Tsallis distribution in this case increases
with increasing $q$.
The Tsallis distribution for quantum statistics has been considered  
in Ref.~\cite{teweldeberhan,uys,turkey1,turkey2}.
%
%
%
%
%
\section{Relation between the Tsallis parameter $q$ and temperature fluctuations}
The parameter $q$ plays a central role in the Tsallis distribution and a physical interpretation
is needed to appreciate its significance.
To this end we follow the analysis 
of Ref.~\cite{wilk} and write the 
Tsallis distribution as a superposition of Boltzmann distributions
\begin{equation}
n^T(E) = \int_0^\infty d\left({1\over T_B}\right)e^{-\left(E-\mu\right)/T_B} 
f\left({1\over T_B}\right)  ,
\end{equation}
where the detailed form of the function $f$ is given in~\cite{wilk}. 
The parameter $T_B$ is the standard temperature as it appears in the Boltzmann distribution.
It is straightforward to show~\cite{wilk} that the average value of $1/T_B$ is given by the 
Tsallis temperature:
\begin{equation}
\left<{1\over T_B}\right> = 
\int_0^\infty  d\left({1\over T_B}\right) \left({1\over T_B}\right)f\left({1\over T_B}\right) = {1\over T}  ,
\end{equation}
while the fluctuation in the temperature is given by the deviation of the 
  Tsallis parameter $q$ 
  from unity:
\begin{equation}
{
\left<\left({1\over T_B}\right)^2\right>
- \left<{1\over T_B}\right>^2 
\over
\left<{1\over T_B}\right>^2 
}
= q - 1  
\end{equation}
which becomes zero in the Boltzmann limit.
The above leads to the interpretation of the Tsallis distribution as a superposition of 
Boltzmann distributions with different temperatures. The average value of these (Boltzmann)
temperatures is the temperature $T$ appearing in the Tsallis distribution. This is the 
interpretation of the Tsallis temperature that we will follow. 
The other parameter in the Tsallis distribution, $q$, describes the spread around the average value
of the (Boltzmann) temperature $T$. For $q=1$ we have an exact Boltzmann distribution, for values of $q$
which deviate from 1, we have a corresponding deviation. 
From this point of view the Tsallis distribution 
describes a distribution of (Boltzmann) temperatures. A deviation from $q=1$
means  that a spread of temperatures  is needed instead of a single value.  
\section{Thermal Fit Details} 
\begin{figure}
\begin{center}
\includegraphics[width=0.8\textwidth,height=10cm]{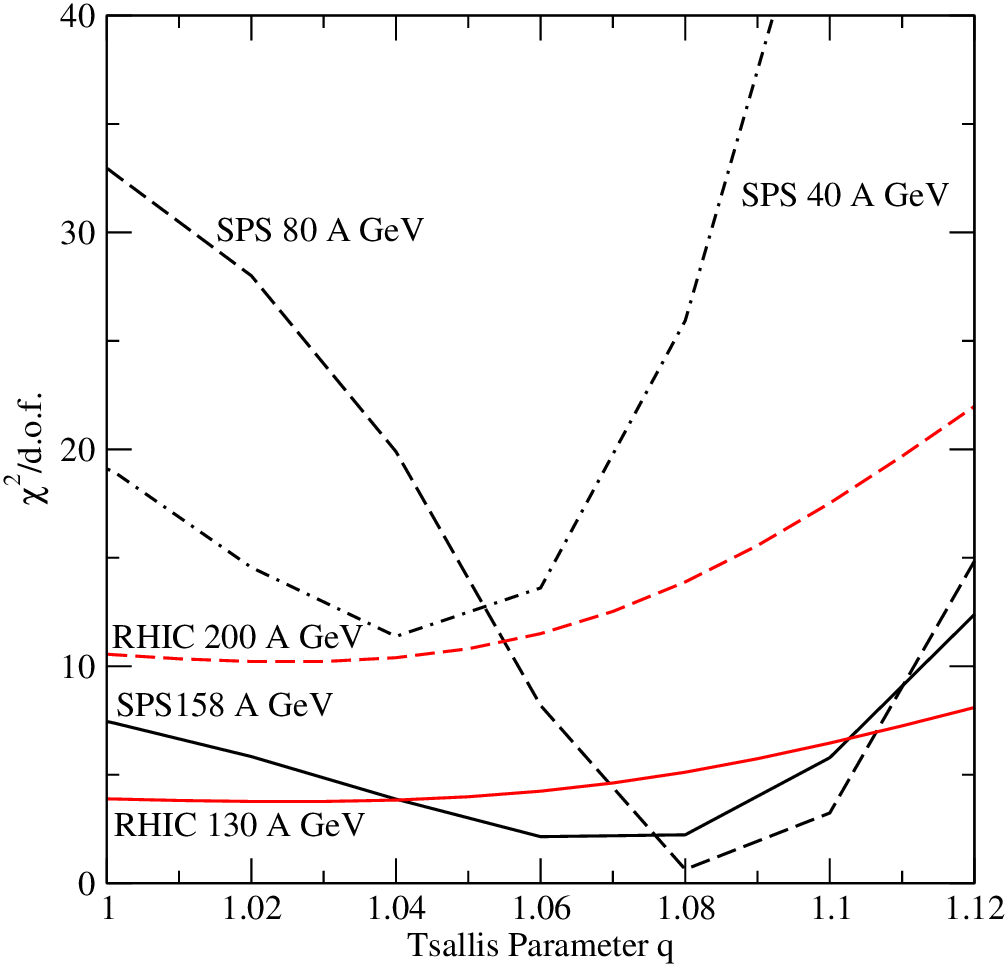}
\caption{The $\chi^2$/d.o.f. of the fits as a function of the Tsallis
parameter $q$.}
\label{chi}
\end{center}
\end{figure}
In order to identify the energy dependence of the deviation from ideal gas 
behaviour, thermal fits were performed on yields measured at the CERN SPS 
in central Pb-Pb collisions at 40 AGeV, 80 AGeV and 158 AGeV (using the same 
data as analyzed in \cite{becattini}) and yields measured at RHIC in 
central Au-Au collisions at $\sqrt{s}$ = 130 AGeV (using the same data as 
analyzed in \cite{WheatonKaneta}) and at $\sqrt{s}$ = 200 AGeV.

In the CERN SPS fits, the thermal parameters $T$, $\mu_B$, $\gamma_S$ 
and $R$ were fit to the data, while $\mu_Q$ and $\mu_S$ were fixed by 
the initial baryon-to-charge ratio and strangeness content in the 
colliding system, respectively.

In the case of the RHIC analysis we again fit $T$, $\mu_B$, $\mu_S$ and $\gamma_S$ 
to the data. The use of mid-rapidity data here led to the relaxing 
of the constraints on $\mu_S$ and $\mu_Q$ typical in analyzes of 4$\pi$ 
data. Instead, $\mu_Q$ was set to zero as justified by the observed 
$\pi^+/\pi^-$ ratio. 
The following expression was used to calculate  primordial particle yields, 
\begin{equation}
n_i^T(E,\gamma_s) = \gamma_s^{\left|S_i\right|}\left[1+\left(q-1\right)\left({E-\mu \over T}\right)\right]^{-1/(q-1)},
\end{equation}
where $\left|S_i\right|$ is the number of valence strange quarks and 
anti-quarks in species $i$. The value $\gamma_s = 1$ obviously corresponds 
to complete strangeness equilibration.
All calculations were done using the THERMUS package~\cite{THERMUS}.
\section{Results and Conclusions}
The most surprising result of our analysis is shown 
in Fig. (\ref{chi}): the quality of the fits, as measured by 
the $\chi^2/\textrm{d.o.f.}$, improves at first as the Tsallis 
parameter $q$ increases. It reaches a minimum value around 
$q\approx$ 1.07 for SPS beam energy of 158 AGeV. This behaviour is repeated 
at other SPS energies with  the minima 
at slightly different values of $q$, i.e. 1.08 for 80 AGeV and about 1.05 at 40 AGeV beam energy.  
This behaviour is not seen at RHIC energies.
Clearly, changes in the Tsallis parameter $q$ have only a small negligible effect on the 
$\chi^2$ values at  RHIC energies, of course, this  still leaves open the possibility
for  $q$ values larger than 1~\cite{Biro:2008km}. 
However on the SPS data the effect is substantial and changes the 
interpretation substantially. One possible  interpretation is that at
SPS energies fluctuations in the freeze-out temperature are
substantial. 

Recently~\cite{biro3} a coalescence model with a Tsallis distribution for quarks
was used  to fit the 
transverse momenta spectra measured at RHIC.  
This fit does not include decays from resonances and therefore cannot be 
compared directly to ours since decays can substantially modify the 
transverse momenta, also the emerging hadrons are not in a Tsallis-type equilibrium gas,
which is an assumption of the present analysis.
The authors
obtained values for the Tsallis parameter $q$ which are remarkably similar
for all particle species considered, i.e. $q\approx 1.2$ which 
cannot be excluded by our analysis. 

\begin{figure}
\begin{center}
\includegraphics[width=0.47\textwidth,height=10cm]{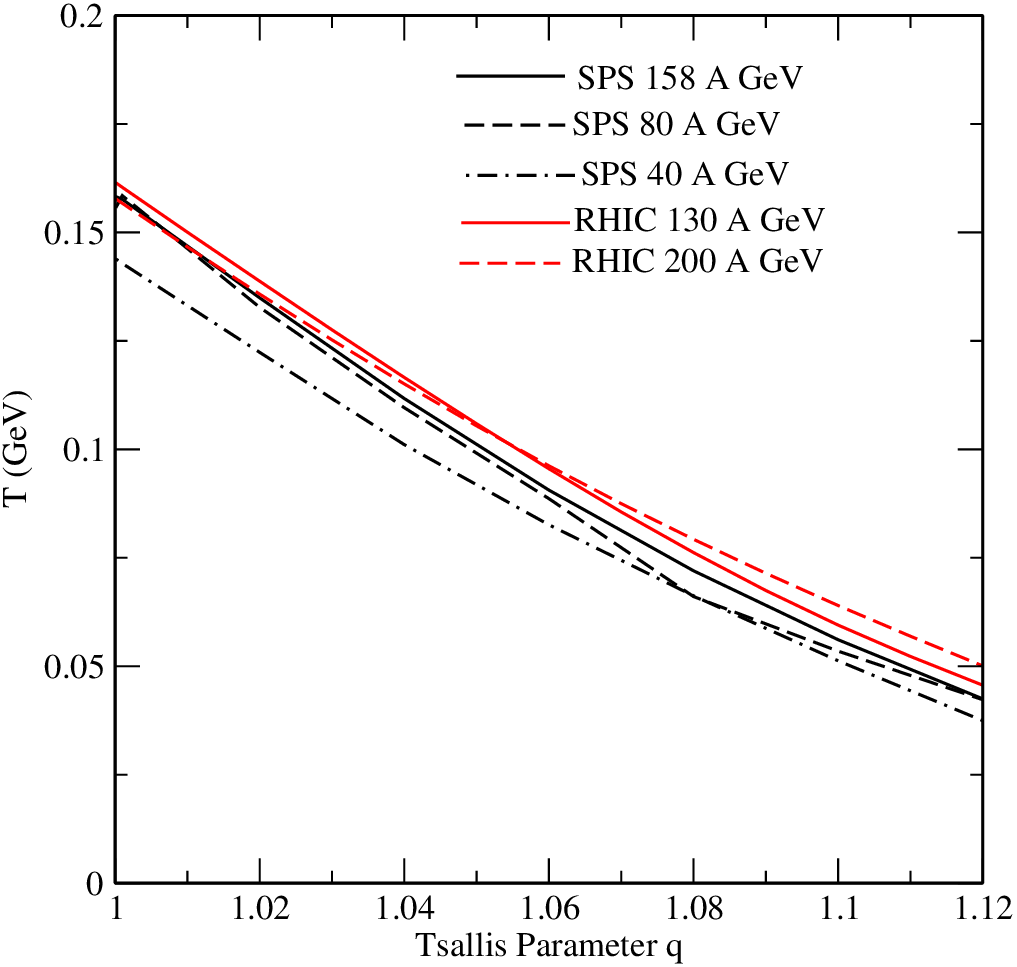}
\includegraphics[width=0.47\textwidth,height=10cm]{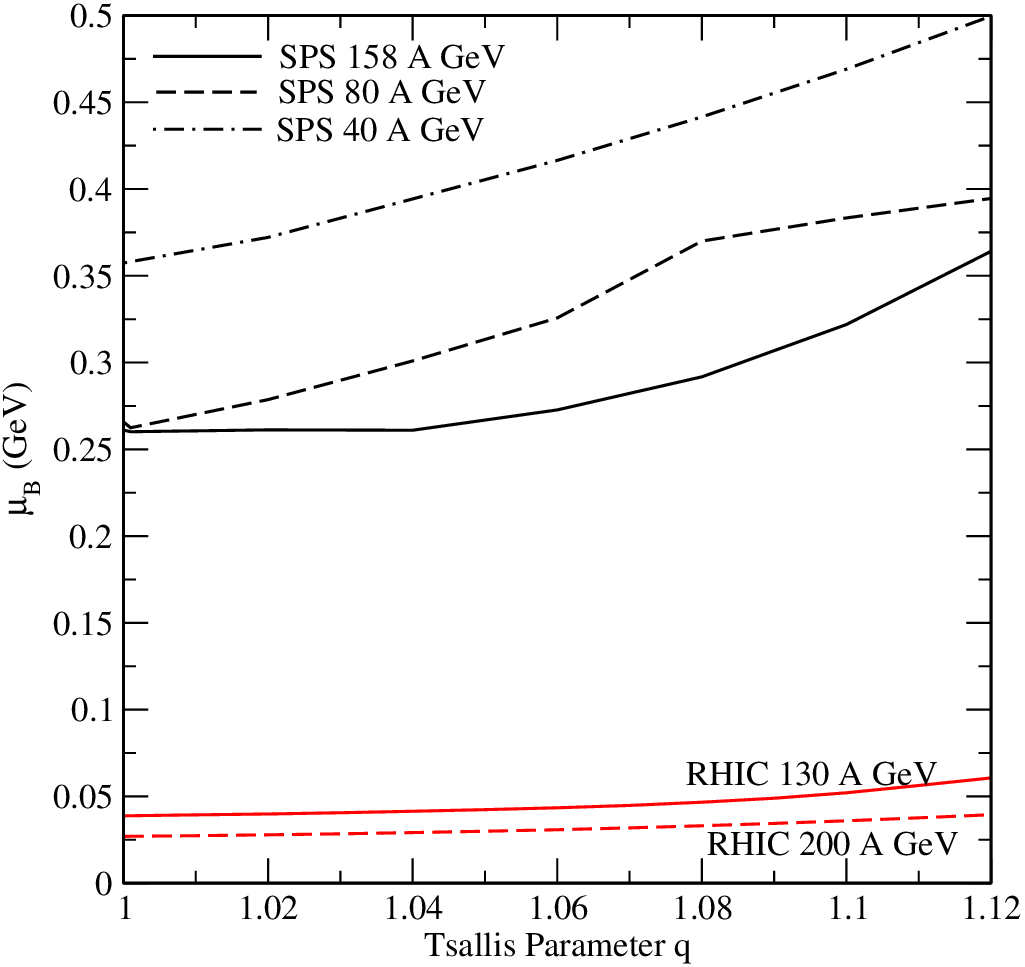}
\caption{The chemical freeze-out temperature (left figure) and the baryon
chemical potential (right figure)  as a function of the Tsallis
parameter $q$.}
\label{temp}
\end{center}
\end{figure}

The freeze-out temperature $T$ decreases, as expected,  with increasing 
values of $q$. This can be understood from the fact that the 
Tsallis distribution is always larger than the Boltzmann one (as long as $q>1$).
Hence, in order to match the same particle yields  one has to adjust $T$ to 
lower values. This is seen at all energies in Fig.~\ref{temp}. 
However, the drop in $T$, turns out to
be quite drastic numerically. In fact, the decrease in particle numbers 
has to be compensated by
increases in all other thermodynamic variables. 
The (modest) increase in the 
baryon chemical potential is shown in Fig.~\ref{temp} on the right hand side.

\begin{figure}
\begin{center}
\includegraphics[width=0.8\textwidth,height=10cm]{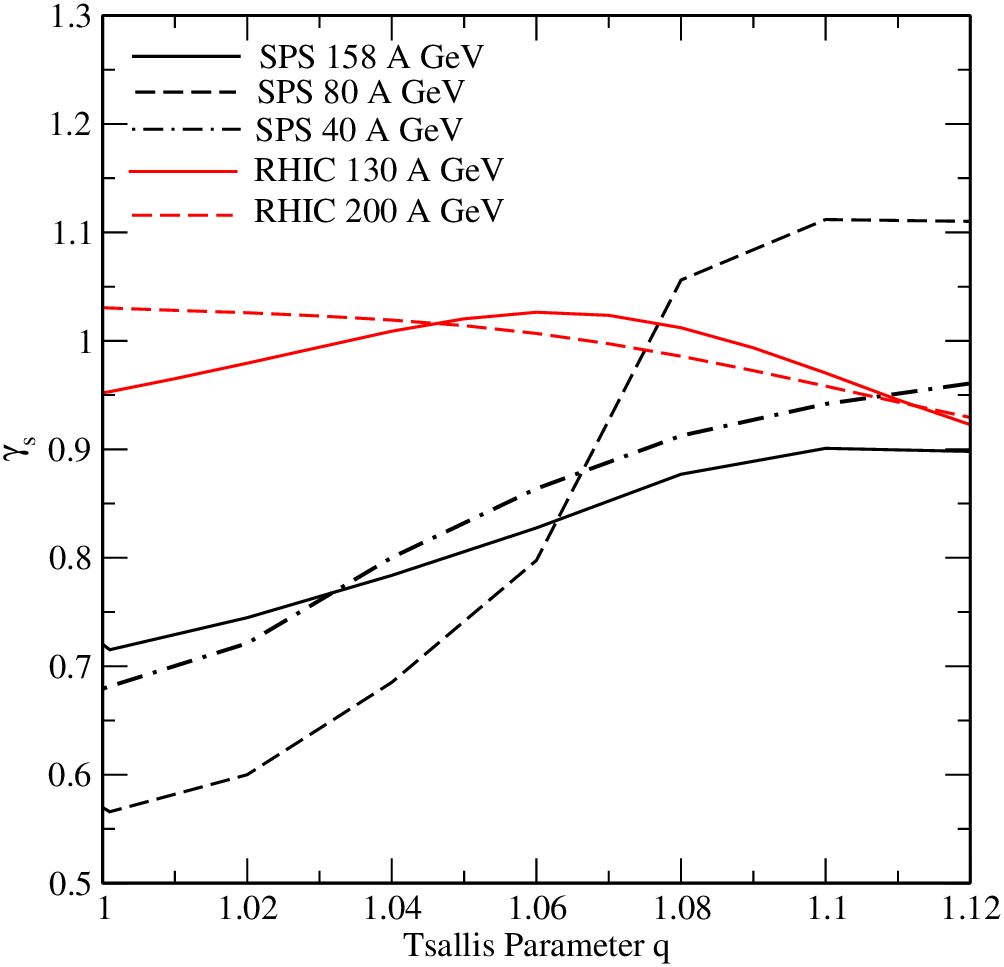}
\caption{The strangeness non-equilibrium factor $\gamma_s$ 
as a function of the Tsallis
parameter $q$.}
\label{gammas}
\end{center}
\end{figure}

The strangeness non-equilibrium factor $\gamma_s$ as 
shown in Fig.~\ref{gammas}. 
It is interesting to note that the 
Tsallis distribution leads to a much better chemical equilibrium than the corresponding 
Boltzmann distribution with $q = 1$. In all cases considered the $\gamma_s$ is very close 
to the chemical equilibrium value of 1.
%
%
%
Clearly, the use of the Tsallis distribution in relativistic 
heavy ion collisions 
calls for a reevaluation  of the understanding gained from 
previous analyses~\cite{wheaton,andronic,becattini}.
\ack
The authors acknowledge support from the Hungary - South Africa 
scientific cooperation programme and Hungarian OTKA grant NK62044.
We acknowledge useful discussions with H.G. Miller, T.S.  Bir\'o, P. V\`an and 
G.G. Barnaf\"oldi.


\section*{References}
\begin{thebibliography}{10}
\bibitem{Fermi50} E.~Fermi, Progress Theor. Phys.,  {\bf 5} (1950) 570.
%
\bibitem{Pomer51} I.Ya. Pomeranchuk, Dokl. Akad. Nauk Ser. Fiz. 78 (1951) 889.
%
\bibitem{heisenberg} W.~Heisenberg, Naturwissenschaften,  {\bf 39} (1952) 69.
%
\bibitem{Haged65} R.~Hagedorn, Nuovo Cimento,  {\bf 35} (1965) 395.
%
\bibitem{wheaton} J.~Cleymans, H. Oeschler, K. Redlich, S. Wheaton, Phys. Rev. C {\bf}73 (2006) 034905.
\bibitem{strange} J.~Cleymans, P. Braun-Munzinger, H. Oeschler and K. Redlich,
Nucl. Phys. A {\bf 697} (2002) 902.
%
\bibitem{andronic} A.~Andronic, P. Braun-Munzinger, J. Stachel, Nucl. Phys.  A {\bf 772} (2006) 167.
\bibitem{becattini} F.~Becattini, J.~Manninen, M.~Ga\'zdzicki Phys. Rev. C {\bf 73} (2006) 044905.
\bibitem{degroot} See e.g. S.R. de Groot, W.A. van Leeuwen, Ch.G. van Weert, {\it Relativistic Kinetic Theory}, North Holland 1980.
\bibitem{tsallis} C.~Tsallis, J. Stat. Phys. {\bf 52} (1988) 479.
\bibitem{tsallis3} C.~Tsallis, R.S.~Mendes, A.R.~Plastina, Physics A {\bf 261} (1998) 534.
\bibitem{biro2} T.S. Bir\'o,  G. Purcsel,  Phys. Rev. Lett. {\bf 95} (2005) 162302.
\bibitem{teweldeberhan} A.M.~Teweldeberhan, A.R.~Plastino, H.G. Miller, Phys. Lett. A {\bf 343} (2005) 71.
\bibitem{uys} A.R.~Plastino, A. Plastino, H.G. Miller, Phys. Lett. A {\bf 343} (2005) 71.
\bibitem{turkey1} F.~Buyukkilic, D. Demirhan, Phys. Lett. A {\bf 181} (1993) 24.
\bibitem{turkey2} F.~Buyukkilic, D. Demirhan, A. Gulec, Phys.  Lett. A {\bf 197} (1995) 209.
\bibitem{wilk} G. Wilk and Z. Wlodarczyk, Phys. Rev. Lett. {\bf 84} (2000) 2770.
\bibitem{THERMUS} S. Wheaton, J. Cleymans, M. Hauer, Computer Physics Communications {\bf 180} (2009) 84.
\bibitem{WheatonKaneta} J. Cleymans, B. K\"{a}mpfer, M. Kaneta, S. Wheaton and N. Xu, Phys. Rev. C {\bf 71} (2005) 054901.
\bibitem{Biro:2008km}
T.S.~Bir\'o, K. \"Urm\"ossy K and G.G. Barnaf\"oldi, J.\ Phys.\ G {\bf 35} 044012 
\bibitem{biro3} T.S.~Bir\'o,  K.~\"Urm\"ossy,  arXiv:0812.2985 [hep-ph].
\end{thebibliography}
\end{document}